\documentclass{PoS}

\usepackage{graphicx}

\title{Dark Matter Signature from the Sky and at Colliders }

\ShortTitle{DM Signature from the Sky  and at Colliders }

\author{ \speaker{Shou-hua Zhu}
\thanks{This project was supported in part by NSF of China under Nos 10775001 and 10635030. }\\
        ITP, Peking University\\
        E-mail: \email{shzhu@pku.edu.cn}}

\abstract{
In this talk, we briefly review our recent investigations on the properties of dark matter (DM) particle.
(1) In 2008 PAMELA released their first results on the positron and antiproton ratios. Stimulated by the new data, we studied \cite{Yin:2008bs} the cosmic ray propagation models and calculated the secondary positron and antiproton spectra. The low energy positron ratio can be consistent with data in the convection propagation model. Above $\sim 10$ GeV PAMELA data shows a clear excess on the positron ratio. However, the secondary antiproton is roughly consistent with data. The positron excess may be a direct evidence of dark matter annihilation or decay. We compare the positron and anti-proton spectra with data by assuming dark matter annihilates or decays into different final states. The PAMELA data actually excludes quark pairs being the main final states, disfavors gauge boson final states. Only in the case of leptonic final states the positron and anti-proton spectra can be explained simultaneously. We also compare the decaying and annihilating dark matter scenarios to account for the PAMELA results and prefer to the decaying dark matter. Finally we consider a decaying neutralino dark matter model in the frame of supersymmetry with R-parity violation. The PAMELA data is well fitted with neutralino mass $600\sim 2000$ GeV and life time $\sim 10^{26}$ seconds. We also demonstrate that neutralino with mass around 2TeV can fit PAMELA and ATIC data simultaneously  \cite{Yin:2008bs}.
We also investigated the corresponding high energy neutrino  \cite{Liu:2008ci} and the synchrotron and inverse Compton radiation \cite{Zhang:2008tb} signatures and found those signature may distinguish different scenarios which can account for the PAMELA observations. (2) The DM sector can be complicated. Motivated by recent observations and theoretical progress, the {\em possible} whole picture of DM sector
emerges, which includes O(TeV) DM and O(GeV) or less light particle. Via the so-called Sommerfeld enhancement effects, the new light particle can fill the gap between the DM thermal relic density and the observed SM particle flux induced by nowadays DM annihilating. The O(TeV) may be detected by LHC. The related O(GeV) or less new particles can be studied by low energy linear colliders (e.g. BESIII) \cite{Zhu:2007zt,Yin:2009mc} or via neutrino telescope \cite{Yin:2009yt}. }

\FullConference{35th International Conference of High Energy Physics - ICHEP2010,\\
		July 22-28, 2010\\
		Paris France}

\begin{document}

\section{Introduction}

Synergy of non-accelerator- and accelerator-based experiments has shaped our knowledge of fundamental particles and interactions. The famous examples were the discoveries of positron and the 'V-particle' in  the
cosmic rays (CRs) in the early stage of particle physics. Since 2008, the PAMELA, ATIC, Fermi, HESS, CDMSII etc. have released their new observations. Some new phenomena were reported. The investigations on their origin and possible connection became extremely necessary. Furthermore whether the resulting underlying picture can be checked by colliders and astrophysical experiments is also worthy of study. This proceeding paper is not the thorough review, instead it contains merely our recent several works \cite{Yin:2008bs,Liu:2008ci,Zhang:2008tb,Zhu:2007zt,Yin:2009mc,Yin:2009yt}.  In section II, we reported our investigation on PAMELA data interpretation. In section III, we describe how to find the new light particle at low-energy collider (BESIII) and via neutrino telescope. Such new light particle at O(GeV) or less is the necessary component to get boost factor via the so-called Sommerfeld enhancement effects.

\section{PAMELA data and leptonically decaying dark matter}

PAMELA is a satellite borne experiment designed to measure CRs in a wide energy range with unprecedented accuracy. Recently, the PAMELA collaboration released
the first data about antiprotons and positrons. Usually it is thought that antiprotons and
positrons are produced when CRs propagate in the Milky Way and
collide with the interstellar medium (ISM). The abundance of these
secondaries can be calculated with relatively high precision.
However, the PAMELA results show an obvious excess in the faction of
$e^+/(e^+ +e^-)$ at energies above $\sim 10$ GeV. Interestingly the
excess keeps to rise up to energy $\sim 100$ GeV. On the other hand,
the spectrum of antiprotons fits the prediction quite well. These
results confirm the previous results by HEAT
and AMS within the error bars.

As a systematic study we investigated firstly the cosmic ray propagation model
carefully so that we can predict the positron and antiproton spectra
to compare with the PAMELA data. Our studies \cite{Yin:2008bs} showed that it is barely to account for
PAMELA observations by adjusting the propagating parameters. Secondly, we adopted a model  independent approach to
constrain the DM annihilation or decay products from the PAMELA
data. We find the PAMELA data actually excludes the annihilation or
decay products being quark pairs, strongly disfavors the gauge
bosons and favors dominant leptonic final states.

Thirdly, we turn to discuss the specific decaying DM model in
supersymmetry scenario.
We find, if the positron excess at PAMELA is indeed of
DM origin, the decaying DM is superior to annihilating DM \cite{Yin:2008bs}.
In SUSY model, the discrete symmetry of
R-parity can be invoked to avoid dangerous baryon-number violation
terms which drive unexpected proton decay. Defined as
$R=(-1)^{2S+3B+L}$, the R-parity of a SM particle is even while
its superpartner is odd. Then the LSP particle is stable.
Since the neutralino LSP can have correct relic density via thermal
production which makes it a suitable DM
candidate.

Although R-parity symmetry is well motivated for SUSY phenomenology,
there is no reason for this symmetry to be exact. One can introduce
some R-parity violation terms in the Lagrangian which make the LSP
decay into SM particles. The general gauge invariant superpotential
of the minimal supersymmetric standard model can be written as
\begin{equation}
W = W_{MSSM}  + \lambda_{ijk} L_i L_j \bar E_k +
\lambda^{'}_{ijk}L_i Q_j \bar D_k + \lambda^{''}_{ijk}\bar U_i
\bar D_j \bar D_k + \mu^{'}_i L_i H_u \label{rvl}
\end{equation}
where i, j, k are generation indices.
In general, we should consider all R-parity violation terms in Eq.
\ref{rvl}. However, they might not appear in theory at the same
time. In
this work, we only consider lepton-number violation term $LL\bar E$
in order to account for the PAMELA positron excess.
The chosen parameters are in Tab.~\ref{DCDR}.

\begin{table}[htb] \begin{center}
\begin{tabular}{|c|c|c|c|c|c|}
\hline \hline DC &  $ \tau (10^{26} s)$
 & $\lambda^{'} (10^{ - 25} )$& DR &  $ \tau (10^{26} s)$
 & $\lambda^{'} (10^{ - 25} )$
  \\
\hline
A&  4.5 & 3.2 &  A  & 3.9 & 3.4 \\
\hline
B&  2.6 & 14.8 &  B  & 2.3 & 15.7 \\
\hline
C&  1.7 & 16.1 &  C  & 1.5 & 17.1 \\
\hline
D&  1.2 & 59.5 &  D  & 1.1 & 62.1 \\
\hline
E&  1.0 & 253.2 &  E  & 0.9 & 266.9 \\
\hline
F&  0.6 & 159.7 &  F  & 0.6 & 159.7 \\
\hline
G&  0.5 & 156.7 &  G  & 0.5 & 156.7 \\
\hline \hline
\end{tabular}
\caption{Life time $ \tau $ in unit of $10^{26} s$ and R-parity
violation parameter $\lambda^{'} $ in unit $10^{ - 25} $ for
different benchmark points. ``DC'' and ``DR'' refer to the two
propagation models we adopted.
}\label{DCDR}
\end{center}
\end{table}

\begin{figure}[h]
\vspace*{-.03in} \centering
\includegraphics[width=3.3in,angle=0]{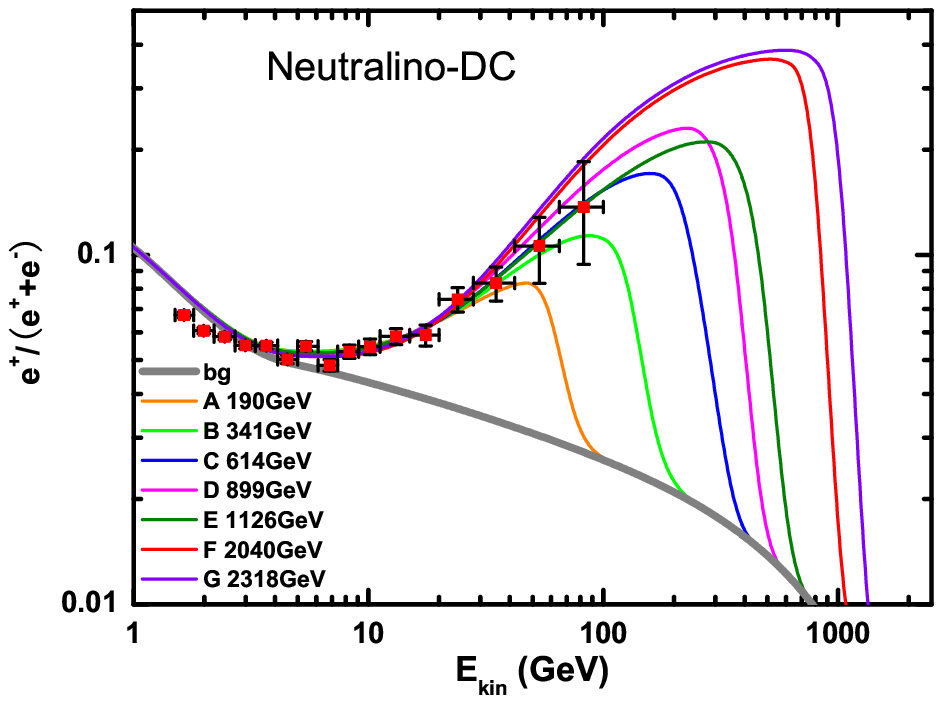}%
\includegraphics[width=3.3in,angle=0]{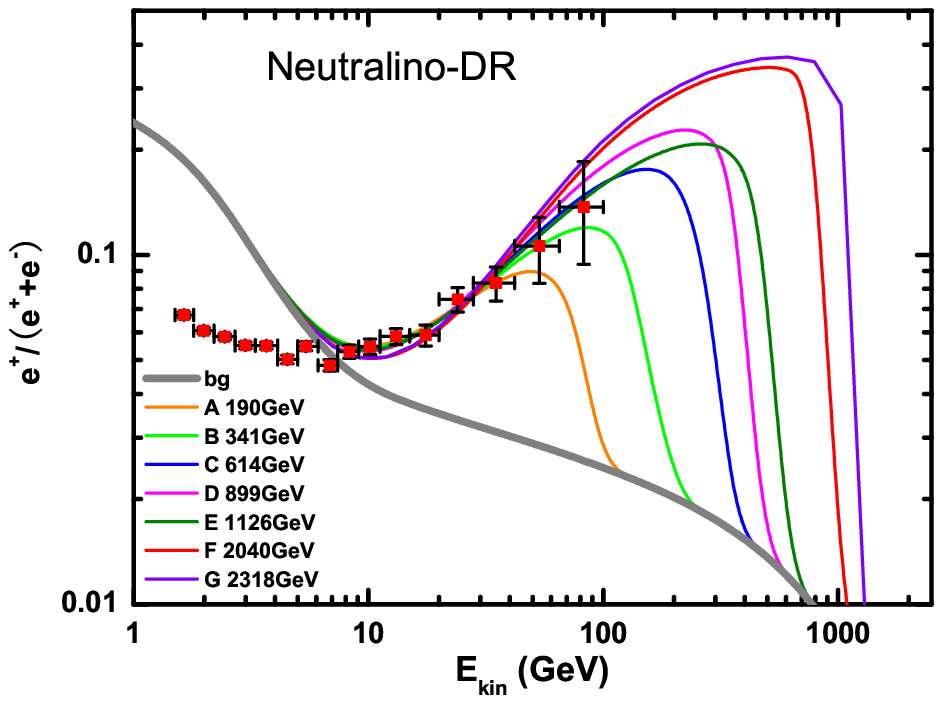}
\vspace*{-.03in} \caption[]{Positron fraction as a
function of energy for different benchmark points. The neutralino
mass of each model is given, while the life time of neutralino in
each model is given in Tab.~\ref{DCDR}. The neutralino mass in each
line is increasing from left to right. } \label{neu}
\end{figure}

The induced positron and antiproton fractions as
function of energy for different benchmark points are shown in Fig.~\ref{neu}.
From the curves we can see clearly that PAMELA observations can be accommodated  in this R-parity-violated model.

\section{New light particle at low-energy collider and neutrino telescope}

In the annihilating DM scenario, there is a long-standing gap between the DM thermal relic density and the observed SM particle flux induced by nowadays DM annihilating.
In the past, provided that DM is O(100) TeV or higher, the weak boson can act as the light particle which induce the required boost factor via the so-called Sommerfeld enhancement effects.
However recent observations implied that the DM is not so heavy, instead only O(TeV). Therefore the {\em possible} whole picture of dark matter sector includes O(TeV) DM and O(GeV) or less new light particle. At the powerful LHC, O(TeV) DM may be detected. Here we only focus on the detection of light particle at low energy collider \cite{Zhu:2007zt,Yin:2009mc} and neutrino telescope \cite{Yin:2009yt}.

 In our studies, the observability of such kind of light particle at BESIII detector is investigated through the processes $ e^ + e^ - \to U\gamma$, followed by $U\to e^+e^-$, $U\to \mu^+\mu^-$ and $U\to \nu\overline{\nu}$. In the invisible channel where light decays into neutrinos, BESIII can measure the coupling of the extra $ U(1)$ down to $ O(10^{- 4}) \sim O(10^{- 5})$ because of the low Standard Model backgrounds. In the visible channel where light particle decays into charged lepton pair, BESIII can only measure the coupling down to $ O(10^{- 3}) \sim O(10^{- 4})$ due to the large irreducible QED backgrounds.

We also investigate the possibility of detecting light long-lived particle (LLP) produced by high energy cosmic ray colliding with atmosphere. The LLP may penetrate the atmosphere and decay into a pair of muons near/in the neutrino telescope ['V-particle' again?]. Such muons can be treated as the detectable signal for neutrino telescope. The dark sector may be complicated, and there may exist more than one light particles, for example the dark gauge boson $A'$ and associated dark Higgs boson $h'$. In our study, we discuss the scenario with $A'$ heavier than $h'$ and $h'$ is treated as LLP. Based on our numerical estimation, we find that the large volume neutrino telescope IceCube has the capacity to observe several tens of di-muon events for favorable parameters if the decay length of LLP can be comparable with the depth of atmosphere. The challenge here is how to suppress the muon backgrounds induced by cosmic rays and atmospheric neutrinos.

\section{Conclusions and discussions }

Single experiment can hardly give us the complete picture of the nature. Synergy of non-accelerator- and accelerator-based experiments as well as the theoretical investigations have shaped and will keep on shaping our knowledge of fundamental particles and interactions.

\end{document}